\documentstyle[aps,prb,twocolumn,epsf,floats]{revtex}
\draft
\begin{document}
\wideabs{

%%%%%%%%%%%%%%%%%%%%%%%%%%%%%%%%%%%%%%%%%%%%%%%%%%%%%%%%%%%%%%%%%%%%%%
\title{Negatively Charged Excitons and Photoluminescence in Asymmetric
Quantum Wells}
%%%%%%%%%%%%%%%%%%%%%%%%%%%%%%%%%%%%%%%%%%%%%%%%%%%%%%%%%%%%%%%%%%%%%%

\author{
   Izabela Szlufarska}
\address{
   Department of Physics,
   University of Tennessee, Knoxville, Tennessee 37996 \\
   Institute of Physics, 
   Wroclaw University of Technology, Wroclaw 50-370, Poland}
\author{
   Arkadiusz W\'ojs}
\address{
   Department of Physics, 
   University of Tennessee, Knoxville, Tennessee 37996 \\
   Institute of Physics, 
   Wroclaw University of Technology, Wroclaw 50-370, Poland}

\author{
   John J. Quinn}
\address{
   Department of Physics, 
   University of Tennessee, Knoxville, Tennessee 37996}

\maketitle
\begin{abstract}
We study photoluminescence (PL) of charged excitons ($X^-$) in narrow
asymmetric quantum wells in high magnetic fields $B$. 
The binding of all $X^-$ states strongly depends on the 
separation $\delta$ of electron and hole layers.
The most sensitive is the ``bright'' singlet, whose binding energy 
decreases quickly with increasing $\delta$ even at relatively small $B$.
As a result, the value of $B$ at which the singlet--triplet crossing 
occurs in the $X^-$ spectrum also depends on $\delta$ and decreases 
from 35~T in a symmetric 10~nm GaAs well to 16~T for $\delta=0.5$~nm.
Since the critical values of $\delta$ at which different $X^-$ states
unbind are surprisingly small compared to the well width, the observation 
of strongly
bound $X^-$ states in an experimental PL spectrum implies virtually no 
layer displacement in the sample.
This casts doubt on the interpretation of PL spectra of 
heterojunctions in terms of $X^-$ recombination.
\end{abstract}
}

\section{Introduction}
%%%%%%%%%%%%%%%%%%%%%%%%%%%%%%%%%%%%%%%%%%%%%%%%%%%%%%%%%%%%%%%%%%%%%%
The optical properties of a quasi-two-dimensional electron gas (2DEG) 
in a high magnetic field $B$ have been widely studied both experimentally
\cite{heiman,kheng,buhmann,shields,finkelstein,hayne,glasberg,munteanu,%
crooker,priest,kim,tischler,jiang,wojtowicz,gravier} and theoretically.
\cite{lampert,stebe,macdonald1,macdonald2,x-dot,palacios,chapman,%
whittaker,x-pl,x-fqhe,x-cf,dzyubenko1,chen,pawel1,brum,pawel2,bilayer}
The 2DEG is usually realized in semiconductor quantum wells (QW) or 
heterojunctions (HJ).
In the QW's, where electrons ($e$) and valence holes ($h$) are confined 
in the same 2D layer, the photoluminescence (PL) spectrum shows emission 
from the radiative states of neutral ($X=e+h$) and charged ($X^-=2e+h$) 
excitons interacting with one another and with free electrons.

The existence of a bound $X^-$ complex was first predicted by Lampert
\cite{lampert} in bulk semiconductors; however, it could not be 
observed experimentally because of the small binding energy $\Delta$. 
It was later shown by Stebe and Ainane\cite{stebe} that the $X^-$ binding
is significantly enhanced in 2D systems.
Indeed, the $X^-$ state with $\Delta$ of about 3~meV has been detected 
by Kheng {\sl et al.}\cite{kheng} in a CdTe QW.
The subsequent extensive experimental\cite{buhmann,shields,finkelstein,%
hayne,glasberg,munteanu,crooker,priest,kim,tischler} and theoretical
\cite{x-dot,palacios,chapman,whittaker,x-pl} studies established that 
the $X^-$ occurs in form of a number of different bound states.
The state observed by Kheng {\sl et al.}\ was the singlet, $X^-_s$, 
whose total electron spin $J$ is zero.
This is the only bound $X^-$ state in the absence of a magnetic field.

MacDonald and Rezayi\cite{macdonald1} showed that the decoupling of 
optically active excitons from electrons in the lowest Landau level (LL) 
due to the ``hidden symmetry''\cite{lerner,dzyubenko2} causes unbinding 
of the $X^-_s$ (and other optically active complexes larger than $X$) 
for $B\rightarrow\infty$.
However, a different bound $X^-$ state exists in this limit.
It is a triplet $X^-_{td}$ with $J=1$ and finite angular momentum 
${\cal L}=-1$.\cite{x-dot} 
Since both the hidden symmetry\cite{palacios,lerner,dzyubenko2} and 
the angular momentum conservation\cite{x-pl,x-fqhe,x-cf,dzyubenko1}
independently forbid recombination of an isolated $X^-_{td}$ in the
lowest LL, its optical lifetime $\tau_{tb}$ in high magnetic fields 
is expected to be long and determined by scattering and/or disorder.

The fact that the $X^-_{td}$ binding energy $\Delta_{td}$ decreases 
with decreasing $B$ implies a singlet--triplet crossing in the $X^-$ 
spectrum at a certain $B$, estimated\cite{whittaker,x-pl} as about 35~T 
for a 10~nm GaAs QW.
Although the PL experiments in high magnetic fields indeed show emission 
from a pair of $X^-$ states,\cite{shields,finkelstein,hayne,glasberg,%
munteanu} neither the crossing has been found\cite{hayne} nor the 
intensity $\tau^{-1}$ of the peak assigned to the triplet state 
increased with increasing $B$ or decreasing electron density.
This apparent discrepancy between theory and experiment has been recently 
resolved by a numerical discovery\cite{x-pl} of yet another $X^-$ state, 
a ``bright'' triplet $X^-_{tb}$.
The $X^-_{tb}$ state has ${\cal L}=0$, $J=1$, large oscillator strength 
$\tau_{tb}^{-1}$, and small binding energy $\Delta_{tb}$, and occurs in 
high magnetic fields in QW's of finite width. 

While the identification of the experimentally observed triplet as the 
$X^-_{tb}$ state explains its small binding energy, the fact that the
more strongly bound $X^-_{td}$ state is not observed confirms its very 
long optical lifetime $\tau_{tb}$.
The reason why $\tau_{tb}$ remains large in the presence of surrounding 
electrons (although the $e$--$X^-$ scattering breaks the ${\cal L}=0$ 
selection rule for an isolated $X^-$) is the short range of $e$--$X^-$ 
repulsion which causes Laughlin $e$--$X^-$ correlations\cite{laughlin,%
halperin} and the effective isolation of all $X^-$ states from the 2DEG.
\cite{x-pl}
These correlations are also responsible for the insensitivity of the 
PL spectra of QW's to the electron density, and for the success of its 
description in terms of the $X^-$ quasiparticles and their single particle 
properties such as binding energy $\Delta$, PL energy $\omega$, or 
oscillator strength $\tau^{-1}$.

The major difficulty in comparing the numerical and experimental data 
is that most experiments are carried out in asymmetrically doped QW's
\cite{shields,finkelstein,hayne,glasberg,munteanu} or HJ's\cite{priest,%
kim} in which an electric field perpendicular to the 2DEG modifies 
confinement and leads to displacement of electron and hole layers.
This displacement or separation between the electron and hole layers,
has been ignored in the existing realistic calculations 
(which take into account the finite widths of the electron and hole layers, 
LL mixing, etc.),\cite{whittaker,x-pl} although from more idealized 
calculations (zero layer widths and no LL mixing)\cite{bilayer} it can
be expected to weaken the $X^-$ binding, possibly in a different manner 
for different $X^-$ states.

In this paper we incorporate the finite electron--hole layer displacement
$\delta$ into the model used earlier\cite{x-pl} to study the $X^-$ states 
in narrow symmetric QW's.
Using exact numerical diagonalization in Haldane's spherical geometry
\cite{haldane1} we examine the dependence of binding energies of all 
different bound $X^-$ states on both magnetic field and the displacement.
In addition to the bright singlet $X^-_s$ (denoted here by $X^-_{sb}$) 
and two triplets, $X^-_{td}$ and $X^-_{tb}$, we identify a dark singlet 
$X^-_{sd}$ with angular momentum ${\cal L}=-2$ which occurs at $\delta>0$, 
in analogy to a known\cite{fox,riva} $D^-$ (charged donor) state at the 
same ${\cal L}$.
We demonstrate that the binding energies of all $X^-$ states strongly 
depend on $\delta$.
Most sensitive is the $X^-_{sb}$ state which unbinds when $\delta$ reaches 
merely $5-10$\% of the QW width (depending on $B$).

Two major conclusions follow from this result:
(i) In the presence of even small layer displacement, the singlet--triplet 
crossing in the $X^-$ spectrum shifts to a considerably lower magnetic 
field (e.g., from 35~T in a symmetric 10~nm GaAs QW to 16~T for 
$\delta=0.5$~nm).
We expect that this could stabilize the hypothetical two-component 
incompressible fluid states involving long-lived $X^-_{td}$ quasiparticles
\cite{x-fqhe,x-cf} and enable its detection in transport experiments.
(ii) The observation of strongly bound $X^-$ states in an experimental 
PL spectrum implies zero or very small layer displacement in the sample
(compared to the QW width). 
While for asymmetrically doped QW's the displacement can be decreased due to 
electron--hole correlations in the direction perpendicular to the QW, 
the interpretation of PL spectra of HJ's in terms of $X^-$ recombination 
is questionable.

\section{Model}
%%%%%%%%%%%%%%%%%%%%%%%%%%%%%%%%%%%%%%%%%%%%%%%%%%%%%%%%%%%%%%%%%%%%%%
In order to preserve a 2D symmetry of a QW in a finite-size calculation,
the electrons and the holes are confined to a Haldane sphere\cite{haldane1} 
of radius $R$.
The magnetic field $B$ normal to the surface is due to a Dirac 
magnetic monopole in the center of the sphere.
The monopole strength $2S$ is defined in the units of elementary flux,
$\phi_0=hc/e$, so that $2S\phi_0=4\pi R^2B$, and the magnetic length is 
$\lambda=R/\sqrt{S}$. 

The single-particle orbitals are called monopole harmonics.\cite{wu,fano} 
They are the eigenstates of angular momentum:
\begin{eqnarray}
\label{eq1}
   L^2 \left|S,l,m\right> &=& \hbar^2 l(l+1) \left|S,l,m\right> 
\nonumber\\
  L_z \left|S,l,m\right> &=& \hbar m \left|S,l,m\right>,
\end{eqnarray}
and their energies,
\begin{equation}
   \varepsilon_{Slm}= \hbar\omega_c
	\left(n+{1\over2}+{n(n+1)\over2S}\right),
\end{equation}
form $(2l+1)$-fold degenerate shells (LL's) labeled by $n=l-S=0$, $1$, 
\dots\ and (in the limit of large $2S$) separated by the cyclotron energy 
$\hbar\omega_c=\hbar eB/\mu c$ (where $\mu$ is the effective electron or 
hole cyclotron mass).

The parameters we used for calculation are appropriate for GaAs/AlGaAs 
QW's of width $w=10$~nm and Al concentration $x=0.33$.
In such structures, mixing between the light- and heavy-hole subbands 
in the valence band is not very strong\cite{whittaker} and both electrons 
and (heavy) holes can be described in the effective mass approximation.
The valence subband mixing enters the model through the dependence of
the effective cyclotron mass of the hole $\mu_h$ on the magnetic field 
(after Cole et al.\cite{cole}).
We omit the Zeeman splitting of electron and hole spin states 
$\left|\sigma\right>$ and only discuss the Coulomb part of the binding 
energy.
While the actual electron and hole $g$-factors depend on the QW width
\cite{snelling} and magnetic field\cite{seck} and on the wavevector $k$ 
(and thus on a particular $X$ or $X^-$ wavefunction\cite{glasberg}), they 
mainly affect the stability of spin-unpolarized complexes\cite{x-pl} and 
much less the splitting of PL peaks for a given polarization of light.
We also neglect mixing between different electron and hole QW subbands
and the (weak\cite{whittaker}) electron--hole correlations in the 
direction perpendicular to the QW (along the $z$ axis).
Instead, we use effective widths of electron and hole layers, $w_e^*$
and $w_h^*$, and their effective displacement $\delta$, which account
both for actual widths and displacement of single-particle wavefunctions 
and for the effects of QW subband mixing and correlations.

Thus, the single-particle states used in our calculation are labeled by 
a composite index $i=[n,m,\sigma]$ and describe an electron or a heavy 
hole with spin projection $\sigma$, whose in-plane quantum numbers are
$n$ and $m$, and the wavefunctions in the $z$-direction are fixed and 
controlled by $w_e^*$, $w_h^*$, and $\delta$.
The electron--hole Hamiltonian can be generally written as
\begin{equation}
\label{eq2}
  H = \sum_{i,\alpha} 
      c_{i\alpha}^\dagger c_{i\alpha} \varepsilon_{i\alpha}
    + \!\!\!\! \sum_{ijkl,\alpha\beta} \!\!\!\!
      c_{i\alpha}^\dagger c_{j\beta}^\dagger c_{k\beta} c_{l\alpha} 
      V^{\alpha\beta}_{ijkl},
\end{equation}
where $c_{i\alpha}^\dagger$ and $c_{i\alpha}$ create and annihilate 
particle $\alpha$ ($e$ or $h$) in state $i$, and $V^{\alpha\beta}_{ijkl}$ 
are the Coulomb matrix elements.
While the 3D Coulomb matrix elements for an arbitrary electron and 
hole density profiles $\varrho(z)$ can be integrated numerically,
\cite{whittaker} we make the following approximation.\cite{x-pl}
For the density functions in the $z$-direction we take $\varrho(z)
\propto\cos^2(\pi z/w^*)$, that is, we replace the actual QW by one with 
infinite walls at the interface and a larger effective width $w^*$.
For a 10~nm GaAs QW's the best fits to the actual wavefunctions are 
obtained for $w_e^*=13.3$~nm and $w_h^*=11.5$~nm. 
The effective 2D interaction
\begin{equation}
   V(r)=\pm\int\! dz\!\int\! dz'
   {\varrho(z)\varrho(z')\over\sqrt{r^2+(z-z')^2}},
\end{equation}
is approximated by $V_d(r)=\pm1/\sqrt{r^2+d^2}$, where the parameter $d$ 
accounts for the finite widths and displacement of the layers.\cite{he}
For the $e$--$e$ repulsion we take $w^*=w_e^*$ and $d=w^*/5$, and for the 
$e$--$h$ attraction $w^*={1\over2}(w_e^*+w_h^*)$ and $d=w^*/5+\delta$. 
The 2D matrix elements of $V_d(r)$ are close to the 3D ones and can be 
evaluated analytically.

The Hamiltonian $H$ is diagonalized numerically for the system of two 
electrons and one hole, in the basis including up to five LL's ($n\le 4$) 
with up to $2S+1=21$ orbitals in the lowest LL.
The eigenstates are labeled by two total angular momentum quantum numbers, 
$L$ and $L_z$, and the total spin of the pair of electrons, $J=0$ or 1.
The conservation of two orbital quantum numbers in a finite Hilbert space 
is the major advantage of using Haldane's spherical geometry to model an 
infinite planar system with the 2D translational symmetry.
The pair of numbers, $L$ and $L_z$, correspond directly to a pair of 
conserved quantities on a plane: total angular momentum projection 
${\cal M}$ and an additional number ${\cal K}$ associated with the 
partial decoupling of the center-of-mass motion in a magnetic field.
\cite{dzyubenko1,avron}
On a sphere, the states within a LL have different $L_z$ and the same 
$L$, and on a plane, they have different ${\cal K}$ and the same 
${\cal L}={\cal M}+{\cal K}$.

The conservation of $L$ (or ${\cal L}$) in the calculation is essential to 
identify of the $X^-$ optical selection rules.\cite{x-pl}
Since the optically active electron--hole pair has $L=0$ (${\cal L}=0$) 
and the electron left behind after the $X^-$ recombination has $l=S$ 
(${\cal L}=0$), only those $X^-$ states at $L=S$ (${\cal L}=0$) are 
radiative (``bright'').
Other (``dark'') states cannot recombine unless the 2D symmetry and the 
resulting angular momentum conservation are broken (e.g., in a 
collision with an impurity or another particle).

The spherical model obviously has some limitations and the most important
one is modification of interactions due to the surface curvature. 
However, if the correlations modeled have a finite (short) range $\xi$ 
that scales with $\lambda$ (as it is for the electron--hole correlations 
that cause binding of the $X^-$ states), $\xi$ can be made small compared 
to $R$ at large $2S$ and the finite-size effects are eliminated in the 
$2S\rightarrow\infty$ limit.

\section{Results and Discussion}
\begin{figure}[t]
\epsfxsize=3.40in
\epsffile{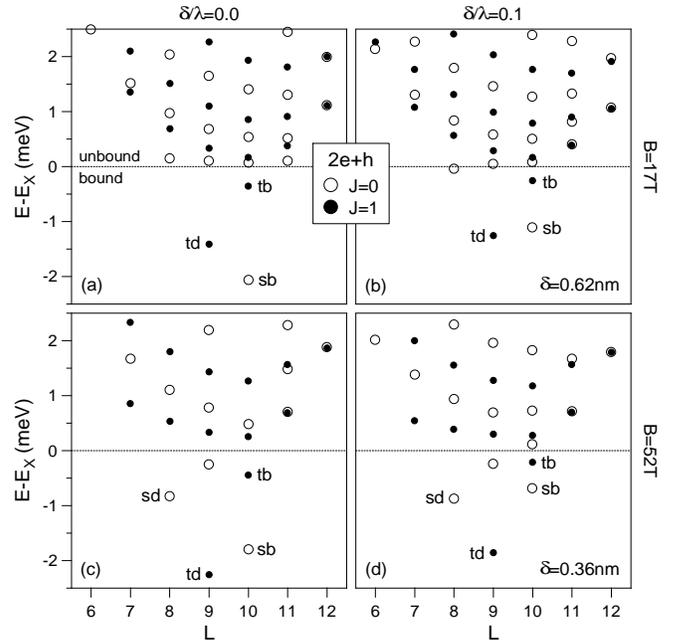}
\caption{
   The energy spectra (energy $E$ vs.\ angular momentum $L$) of the 
   $2e+h$ system on a Haldane sphere with $2S=20$.
   Open and full circles mark singlet and triplet states, respectively.
   The magnetic field is $B=17$~T (ab) and $B=52$~T (cd).
   The layer displacement is $\delta/\lambda=0$ (ac) and 
   $\delta/\lambda=0.1$ (bd).
   $\lambda$ is the magnetic length.}
\label{fig1}
\end{figure}

%%%%%%%%%%%%%%%%%%%%%%%%%%%%%%%%%%%%%%%%%%%%%%%%%%%%%%%%%%%%%%%%%%%%%%
The $2e+h$ low energy spectra for two different values of $B=17$ and 
52~T, and at $\delta/\lambda=0$ and 0.1 are shown in Fig.~\ref{fig1}.
The calculation was carried out for $2S=20$ and including five LL's 
($n\le4$).
We have checked\cite{x-pl} that these numbers are sufficient to obtain
quantitatively meaningful results.
The energy $E$ is measured from the exciton energy $E_X$, so that for 
the bound $X^-$ states below the dotted lines, the vertical axes show 
the negative of their binding energy, $-\Delta=E-E_X$.
Singlet ($J=0$) and triplet ($J=1$) states are marked with open and full
dots, respectively.
The energy is plotted as a function of total angular momentum and each
data point represents a degenerate $L$-multiplet.

The states of particular interest are the bound states with the largest 
$\Delta$ and/or the bright states at $L=S$.
Depending on $B$ and $\delta$, we identify all or some of the following
bound $X^-$ states in the spectrum: 
bright singlet $X^-_{sb}$ at $L=S$ (${\cal L}=0$), 
dark singlet $X^-_{sd}$ at $L=S-2$ (${\cal L}=-2$), 
bright triplet $X^-_{tb}$ at $L=S$ (${\cal L}=0$), 
and dark triplet $X^-_{td}$ at $L=S-1$ (${\cal L}=-1$).
As shown in Fig.~\ref{fig1}(ac), in the absence of layer displacement
the $X^-_{sb}$ is the ground state at the lower magnetic field of 
$B=17$~T, but at a higher magnetic field of $B=52$~T it is $X^-_{td}$ 
that has the lowest energy.
Another bright state $X^-_{tb}$ occurs in the spectrum, but it has higher 
energy than $X^-_{sb}$ or $X^-_{td}$ at all fields.
There is also a dark $X^-_{sd}$ state that becomes bound at a sufficiently
large $B$, but it is not expected to affect the PL spectrum because it
is neither radiative nor strongly bound at any $B$.
The situation is dramatically different when a finite layer displacement 
is included in Fig.~\ref{fig1}(bd).
For $\delta=0.1$~$\lambda$, the binding energies of all $X^-$ states are 
significantly reduced.
The most affected is the bright singlet $X^-_{sb}$ which is no longer 
the ground state even at a relatively low magnetic field of $B=17$~T.
It is quite remarkable that a displacement as small as $\delta=0.62$~nm 
(at $B=17$~T) or $\delta=0.36$~nm (at $B=52$~T), that is only a few 
percent of the QW width of $w=10$~nm and certainly could be expected 
in asymmetric QW's, causes such reconstruction of the $X^-$ spectrum.
The ground state transition from the bright singlet to the dark triplet 
induced at lower $B$ is similar to that caused by a magnetic field at 
$\delta=0$.\cite{whittaker,x-pl}

\begin{figure}[t]
\epsfxsize=3.40in
\epsffile{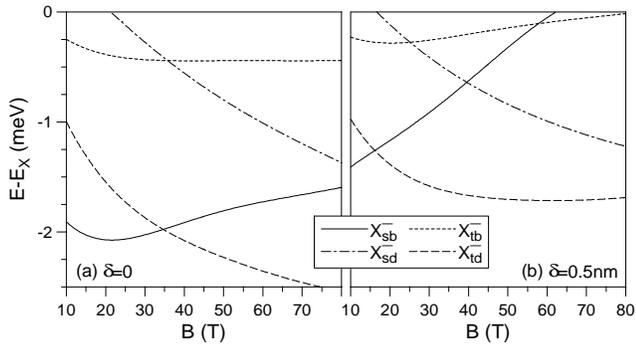}
\caption{The $X^-$ binding energies $E$ calculated on a Haldane sphere
 with the LL degeneracy $2S+1=21$, plotted as a function of the magnetic
field $B$. The parameters are appropriate for a 10~nm GaAs quantum well.
The layer displacement is $\delta=0$ (a) and $\delta=0.5$~nm (b).
}
\label{fig2}
\end{figure}

The effect of the layer displacement on the dependence of the $X^-$ 
binding energies on the magnetic field is shown in Fig.~\ref{fig2}. 
At $\delta=0$, the binding energies of the two bright states remain 
almost constant over a wide range of $B$, in contrast to the two dark 
states which quickly gain binding energy when $B$ increases.
As found in the previous studies,\cite{whittaker,x-pl} this different 
$\Delta(B)$ dependence results in a singlet--triplet ground state 
transition at $B\approx35$~T.
At a small displacement of $\delta=0.5$~nm, the binding energy of the 
bright singlet $X^-_{sb}$ decreases rather quickly as a function of $B$,
more so than the binding energies of other $X^-$ states.
As a result, the singlet--triplet transition occurs at a much lower
magnetic field of $B\approx16$~T and the bright singlet unbinds completely
at $B$ larger than about 60~T.
Actually, neither bright state is strongly bound at $B>60$~T, while the 
binding energies of both dark states remain fairly large (e.g., 
$\Delta_{sd}=1.0$~meV and $\Delta_{td}=1.7$~meV at $B=60$~T).

\begin{figure}[t]
\epsfxsize=3.40in
\epsffile{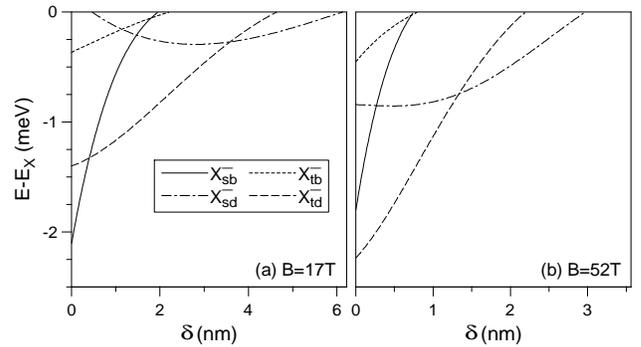}
\caption{The $X^-$ binding energies vs. the displacement of the electron
and hole layers in a 10~nm asymmetric quantum well. The magnetic field is
$B=17$~T (a) and $B=52$~T (b). $E_X$ is the exciton energy.
}
\label{fig3}
\end{figure}

To illustrate the effect of the layer displacement on the $X^-$ states
most clearly, in Fig.~\ref{fig3} we plot the $X^-$ binding energies as 
a function of $\delta$ for two values of the magnetic field.
In both frames, $\delta$ goes from 0 to 0.1~$\lambda$ (where $\lambda=
6.2$ and 3.6~nm at $B=17$ and 52~T, respectively).
For $B=17$~T the ground state transition from $X^-_{sb}$ to $X^-_{td}$ 
occurs at $\delta=0.4$~nm and for $B=52$~T the $X^-_{td}$ is the ground 
state at all displacements.
It is clear that the displacement has more effect on the binding energy 
of $X^-_{sb}$ than on the binding energy of the next most strongly bound 
state, $X^-_{td}$.
This can be understood by noting that the $X^-_{sb}$ complex has smaller 
$|{\cal L}|$ and thus smaller average electron--hole distance 
$\left<r_{eh}\right>$, and that the effect of a finite $\delta$ in $V_d(r)$ 
decreases as $r$ increases.

Let us point out that the binding energies obtained here are rather 
sensitive not only to $B$ or $\delta$, but also to other details of our 
model, including some of its simplifications or approximations.
For example, a slightly different approximation used here to calculate
the $e$--$h$ Coulomb matrix elements at $\delta=0$ resulted in smaller 
binding energies compared to Ref.~\onlinecite{x-pl} (although the 
difference in $\Delta$ appears to be similar for all $X^-$ states, and
the singlet--triplet crossing is obtained at the same $B$, which means 
that the difference between the models affects $E_X$ rather than $E_{X^-}$).
Whittaker and Shields\cite{whittaker} showed that even in narrow QW's 
the inclusion of higher QW subbands and electron--hole correlations in 
the $z$-direction enhances somewhat the $X^-$ binding, specially that 
of the $X^-_{sb}$ state.
Based on their calculation, one can expect that our values obtained 
in the lowest subband approximation are underestimated by up to 0.5~meV, 
depending on $B$ and the particular $X^-$ state.
Despite the difficulty with obtaining definite values of $\Delta$, two 
conclusions arising from our calculation seem quite important and at 
the same time independent of the approximations made.

(i) 
Even a small displacement of electron and hole wavefunctions in the 
$z$-direction shifts the singlet--triplet transition to a considerably
lower value of the magnetic field.
Therefore, the theoretical value of $B\approx35$~T for the crossing 
in a 10~nm well must 
be understood as the upper estimate, and in an experimental sample the
crossing  can occur at any smaller value.
This effect broadens the range of magnetic fields in which the $X^-_{td}$'s 
together with electrons are both most stable and long-lived quasiparticles
in the electron--hole system.
It thus seems that the proposed\cite{x-fqhe,x-cf} incompressible fluid 
states of $X^-_{td}$'s and electrons could be observed more easily in 
slightly asymmetric QW's.

(ii)
The binding energies of both bright $X^-$ states are strongly sensitive 
to the layer displacement.
Therefore, the recombination from strongly bound $X^-$ states observed 
in an experimental PL spectrum implies zero or very small displacement 
in the sample (compared to the QW width). 
The parameter $\delta$ used in our model describes displacement of electron 
and hole wavefunctions in the $z$-direction within a particular bound $X$ 
or $X^-$ state and must be distinguished from the bare displacement 
$\delta_0$ of single-electron and single-hole wavefunction due to an 
external electric field (e.g., caused by a charged doped layer).
It is therefore possible that even in strongly asymmetric QW's, 
the electron--hole correlations in the $z$-direction (which favor small 
displacement) dominate the effect of external electric field (which
causes displacement), and the resulting $\delta$ is much smaller than
$\delta_0$.
If correct, this picture of symmetry (partially) restored by correlations 
explains the success of ``symmetric models''\cite{whittaker,x-pl} to 
describe a wide class of symmetric and asymmetric QW's (and invalidates 
the use of the lowest subband approximation with $\delta_0$ taken for 
unbound particles).
However, it does not seems possible that any $X^-$ states should form
in HJ's where the electrons are confined in a narrow 2D layer and the 
holes remain outside of this layer.
Consequently, the interpretation of multiplets in the PL spectra of HJ's 
in terms of $X$ and $X^-$ recombination seems questionable.
A recent alternative interpretation\cite{bilayer} involves coupling of
a distant hole to (Laughlin) charge excitations of the 2DEG and formation
of bound and radiative (fractionally charged) excitonic complexes of a 
different type.

\section{Conclusion}
%%%%%%%%%%%%%%%%%%%%%%%%%%%%%%%%%%%%%%%%%%%%%%%%%%%%%%%%%%%%%%%%%%%%%%
Using exact numerical diagonalization in Haldane's spherical geometry,
we have studied the effect of the displacement $\delta$ of electron and 
hole layers on the binding energies of the $X^-$ states formed in narrow 
asymmetric QW's in high magnetic fields $B$.
Depending on $B$ and $\delta$, different bound $X^-$ states were 
identified in the $2e+h$ spectrum: bright singlet $X^-_{sb}$, dark singlet 
$X^-_{sd}$, bright triplet $X^-_{tb}$, and dark triplet $X^-_{td}$.
The binding energies of all $X^-$ states quickly decrease as a function
of $\delta$.
The most sensitive is the strongly bound $X^-_{sb}$ state, and
even at displacements very small compared to the QW width, the 
magnetic field induced transition from this bright ground state to 
the dark $X^-_{td}$ ground state occurs at significantly lower values 
of $B$.
The critical displacement for which the bright $X^-$ states unbind is 
only $5-10$\% of the QW width (depending on $B$).
Therefore, detection of the $X^-$ recombination in an experimental PL 
spectrum implies virtually no displacement of electron and hole layers 
(within the observed $X^-$ states). 
While in asymmetric QW's small values of $\delta$ can result from 
electron--hole correlations, the interpretation of the PL spectra of 
HJ's in terms of $X^-$'s is questionable. 

\section*{Acknowledgment}
%%%%%%%%%%%%%%%%%%%%%%%%%%%%%%%%%%%%%%%%%%%%%%%%%%%%%%%%%%%%%%%%%%%%%%
The authors wish to thank S.A. Crooker (LANL Los Alamos) for helpful discussions.
The authors acknowledge partial support of Grant DE-FG02-97ER45657 from 
the Materials Science Program -- Basic Energy Sciences of the US 
Department of Energy.
I.S. and A.W. acknowledge partial support of Grant 2P03B11118 from the 
Polish Sci.\ Comm.

\end{document}